\def\be{\begin{equation}}
\def\ee{\end{equation}}

\documentclass[12pt,a4paper,epsfig]{iopart}
\usepackage{epsfig}
\usepackage{graphicx}
\usepackage[below]{placeins} 
\usepackage{textcomp}
\usepackage{color}
\usepackage{amssymb}

\newcommand{\labs} {\left\vert}
\newcommand{\rabs} {\right\vert}

\newcommand{\lab} {\left\langle}
\newcommand{\rab} {\right\rangle}

\makeatletter
\renewcommand{\@fnsymbol}[1]{%
   \ifcase#1\or\textasteriskcentered\or\textsection\or\textdagger
   \or\textdaggerdbl\or\textparagraph\or\textbardbl
   \or\textasteriskcentered\textasteriskcentered
   \or\textdagger\textdagger\or\textdaggerdbl\textdaggerdbl
   \else\@ctrerr\fi}
\makeatother

\begin{document}
\title[Limited robustness of edge magnetism]
      {Limited robustness of edge magnetism in zigzag graphene nanoribbons with
      electrodes}
\author{S. Krompiewski}%
\address{Institute of Molecular Physics, Polish Academy of
Sciences, ul.~M.~Smoluchowskiego 17, 60179 Pozna{\'n}, Poland
}%
\date{\today}


 \begin{abstract}

It is shown that apart from well-known factors, like temperature,
substrate, and edge reconstruction effects, also the presence of
external contacts is destructive for the formation of magnetic
moments at the edges of graphene nanoribbons. The edge magnetism
gradually decreases when graphene/electrode interfaces become more
and more transparent for electrons. In addition to the
graphene/electrode coupling strength, also the aspect ratio
parameter, i.e. a width/length ratio of the graphene nanoribbon,
is crucial for the suppression of edge magnetism.

The present theory uses a tight-binding method, based on the mean-field Hubbard
Hamiltonian for $\pi$ electrons, and the Green's function technique within
the Landauer-B{\"u}ttiker approach.

\end{abstract}
\maketitle

\section{Introduction}

The problem of edge states in carbon nanoribbons is an important
issue which has been intensively studied for almost 20 years now,
i.e since the edge states were first theoretically predicted
\cite{Nakada_PRB1996}. Experimental confirmations of this concept
came a few years later (\cite{Klusek_ASS2000} and
\cite{Kobayashi_PRB2005}). Since then, a lot of theoretical papers
have been devoted to the edge state problem. Magnetic edge states
were examined, e.g. in Refs \cite{Wimmer_prl2008}-
\cite{Gosalbez_ssc2012}, where fundamental for spintronics
phenomena - such as spin transport, giant magnetoresistance and
Coulomb blockade effects - were studied. Recently the problem has
been given an additional impetus due to new challenges related
with a rich class of graphene-like materials. Some of them, in
contrast to graphene, possess a significant spin-orbit coupling
and are of interest as potential quantum spin Hall systems and/or
topological insulators
(\cite{Xu_nanoscale2012}-\cite{Rachel_prb2014}).

 The edge states become of particular
 importance in nanostructures because then a great fraction of atoms lies at the edges, what strongly influences electronic - and thereby chemical,
 electrical and probably magnetic - properties. The latter will be focused on in this study. It is well-known that at zigzag type fragments of graphene's
 edges the electronic states are of localized nature, with the exponentially decaying amplitude as the interior is approached. These states have very flat
 energy spectra in the vicinity of the charge neutrality point, hence their density of states may be high enough to satisfy the so-called Stoner criterion
 for the appearance of magnetism. The problem is still open, and there is a lively debate on it. On the one hand there are sceptic opinions, stressing that
 a number of factors can destroy the edge state magnetism, including the temperature, reconstruction (or closure) of edge atoms, as well as passivation thereof
 \cite{Kustmann_PRB2011}.
  On the other hand however there are already experimental results obtained by STM and STS (scanning tunneling microscopy and spectroscopy) methods which provide results supporting the view that
  spin-degeneracy of edge states may be lifted \cite{Tao_NatPhys11}. Complexity of the problem has recently been convincingly presented in Ref. \cite{Li_PRL2013},
  where it is pointed out that whether or not edge magnetic
 moments can exist depends also on the substrate material they are in contact with.

In this study, it is shown that $\pi $-electron edge states can
even be affected if the edge atoms are relatively far away (up to
a few tens of angstroms) from the non-local disturbance due to
contacts. This is in contrast to the hitherto  known situations
corresponding to the edge reconstruction \cite{Koskinen_PRB2009},
edge atom closure and hydrogenation \cite{Kustmann_PRB2011}, as
well as the temperature and substrate effects \cite{Li_PRL2013}.
Indeed, it results from the present findings that carbon edge
atoms distant from the external electrodes may lose their magnetic
moments, unless the GNR is weakly coupled to the contacts. This
means that the promising concepts to use GNRs as spintronic
devices (see \cite{Son_Nat2006,Kin_NatNano2008,Yazyev_RPR2010}),
might be realized provided that electrodes are properly selected.
In particular contacts ensuring the formation of high resistive
interfaces with the graphene nanoribbons (GNR) - possibly
tunneling junctions - should be applied.

\section{Methodology and Modeling}

The present approach is based on the tight-binding Hubbard model in the mean-field approximation.

\begin{eqnarray} \label{H}
H &=& - \sum \limits_{ i ,j, \sigma } t_{i,j}\labs i,
\sigma \rab \lab \sigma, j \rabs +\frac{1}{2}\sum \limits_{i, \sigma}\Delta_{i, \sigma} \labs i, \sigma \rab \lab \sigma, i \rabs , \\
\Delta_{i,\sigma} &=& U(n_{i,\sigma}-n_{i,-\sigma}),
\end{eqnarray}

with the nearest neighbor hopping parameter $t=t_{i,j}$,
intra-atomic Coulomb repulsion U and the $\sigma$-spin occupation
number $n_{i,\sigma}$ at the lattice site \textrm{i}. Well
established values of the hopping parameter for graphene ($t_G$)
range between 2.7 eV and 3 eV. As concerns the U parameter, it is
still under debate (\textit{cf.} U equal to 0.5 eV in
\cite{Nair_NatComm013} and U=1.2 $t_G$ in \cite{Yazyev_RPR2010}).
In this paper a moderate value of $U/t_G = 0.6$ is used.

On the one hand the occupation number in the case of non-contacted
(free standing) GNRs can be determined, after having diagonalized
the Hamiltonian (\ref{H}), $(H-E)\labs u_E \rab=0$, as

\begin{equation}  \label{n1}
 n_{i,\sigma}= \sum \limits_{E<E_F} \labs u_E^{i,\sigma} \rabs ^2
\end{equation}

On the other hand in the presence of external contacts, the occupation number can be found, applying the Green's function technique, in terms of  self-energies $\Sigma_1$ and $\Sigma_2$
for source and drain electrodes, respectively.
The corresponding equations read:

\begin{eqnarray} \label{n2}
 n_i &=& -\frac{1}{\pi} \int_{-\infty}^{E_F} dE \; \; Im {\cal G}_{i,i}(E), \nonumber \\
 \cal G &=& (\hat 1 E-H-\Sigma_1-\Sigma_2)^{-1}, \nonumber \\
 \Sigma_{1,2} &=& \hat t_c \; g_{1,2} \; \hat t_c^\dagger.
\end{eqnarray}

   \begin{center}
  \begin{figure}[t]
\includegraphics[width=9cm,height=5cm,angle=0]{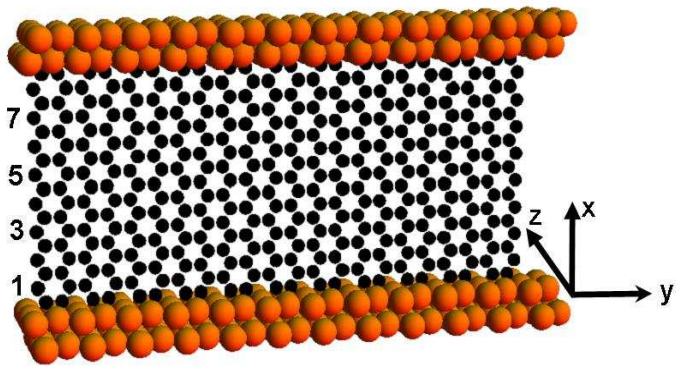}
 \caption{\small (Color online) Perspective view of the 3-dimensional setup
 composed of the (11, 9) GNR (small spheres in the x-y page plane), and the electrodes infinite
in the y-z plane, and semi-infinite in the x direction (only atoms close to the
interface - larger spheres - are shown). Carbon atoms are
enumerated column-wise, from the bottom upwards.} \label{fig1}
\end{figure}
\end{center}

Above, the self-energies are defined as products of the
electrode/GNR hopping parameter matrices ($\hat t_c$) and
corresponding surface Green's functions (\textit{g}) of the
electrodes.
The contacts are modeled by surface Green's functions of a close
packed semi-infinite face centered cubic lattice, fcc(111), which
for the single-band model used here are known analytically (see
\cite{Todorov-JPCM93,Krompiewski_prb2009, SK-Nano12} for details).
No periodic boundary conditions have been employed.

Now it is possible to compute the following physical properties of interest here, \textit{i.e.}
local magnetic moments (m), conductance (G), transmission (T) and shot noise Fano factor:

\begin{eqnarray} \label{GMR}
 m &=& n_{i,\uparrow}-n_{i,\downarrow},\nonumber \\
 G  &=& \frac{e^2}{h} Tr(T), \; \; \; T = \Gamma_1 {\cal G} \Gamma_2 {\cal G}^\dagger, \nonumber \\
 \Gamma_{1,2} &=&
i(\Sigma_{1,2}-\Sigma_{1,2}^{\dagger}), \nonumber \\
 F & = &  1-Tr(T^2)/Tr (T).
\end{eqnarray}

A representative setup considered here is presented in
Fig.~\ref{fig1}.

The GNRs are characterized by two numbers ($N_a$, $N_z$) related
with the number of unit cells in the armchair and zigzag
directions, respectively. Thus, the number of carbon atoms along
the edge (interface) boundary is equal to $N_z$-1 (2 $N_a$). The
undercoordinated edge atoms are assumed to be passivated by
hydrogen, so there are no $\sigma$-type dangling bonds.


\section{Results}

The most important question we focus on is the influence of the
contacts on the edge magnetism, i.e. magnetic moments of the
carbon atoms along the outermost zigzag lines. In order to
elucidate this point a number of GNRs has been studied. It turns
out that apart from the obvious relevance of the interface
coupling, modeled by a hopping integral strength between
GNR/contact atoms $t_c$, also the aspect ratio  of the GNR defined
as A=width/length is of importance. So in the following, edge atom
magnetic profiles are presented both for non-contacted ($t_c=0$)
as well as contacted GNRs of various A parameters.

\begin{center}
  \begin{figure}[h]
\includegraphics[width=9cm,height=5cm,angle=0]{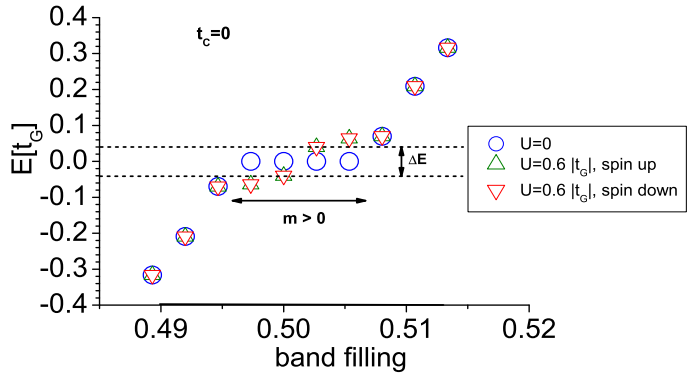}
 \caption{\small (Color online) Energy spectrum of the (11, 9) free standing GNR close to the half band
 filling for non-magnetic and magnetic cases (sphere and triangle markers, respectively).
 Away from the charge neutrality point, all the markers coincide, meaning disappearance of the
 magnetism.} \label{E_tc0}
   \end{figure}
   \end{center}

A typical energy spectrum around the charge neutrality point for a GNR is presented in Fig.~\ref{E_tc0}.

 It is consistent with the next figure (Fig.~\ref{GiF_tc.15}) representing the
conductance spectrum, with contacts relatively
weakly coupled to the GNR ($t_c/t_G=0.15$).

 \begin{center}
  \begin{figure}[b]
\includegraphics[width=8cm,height=5cm,angle=0]{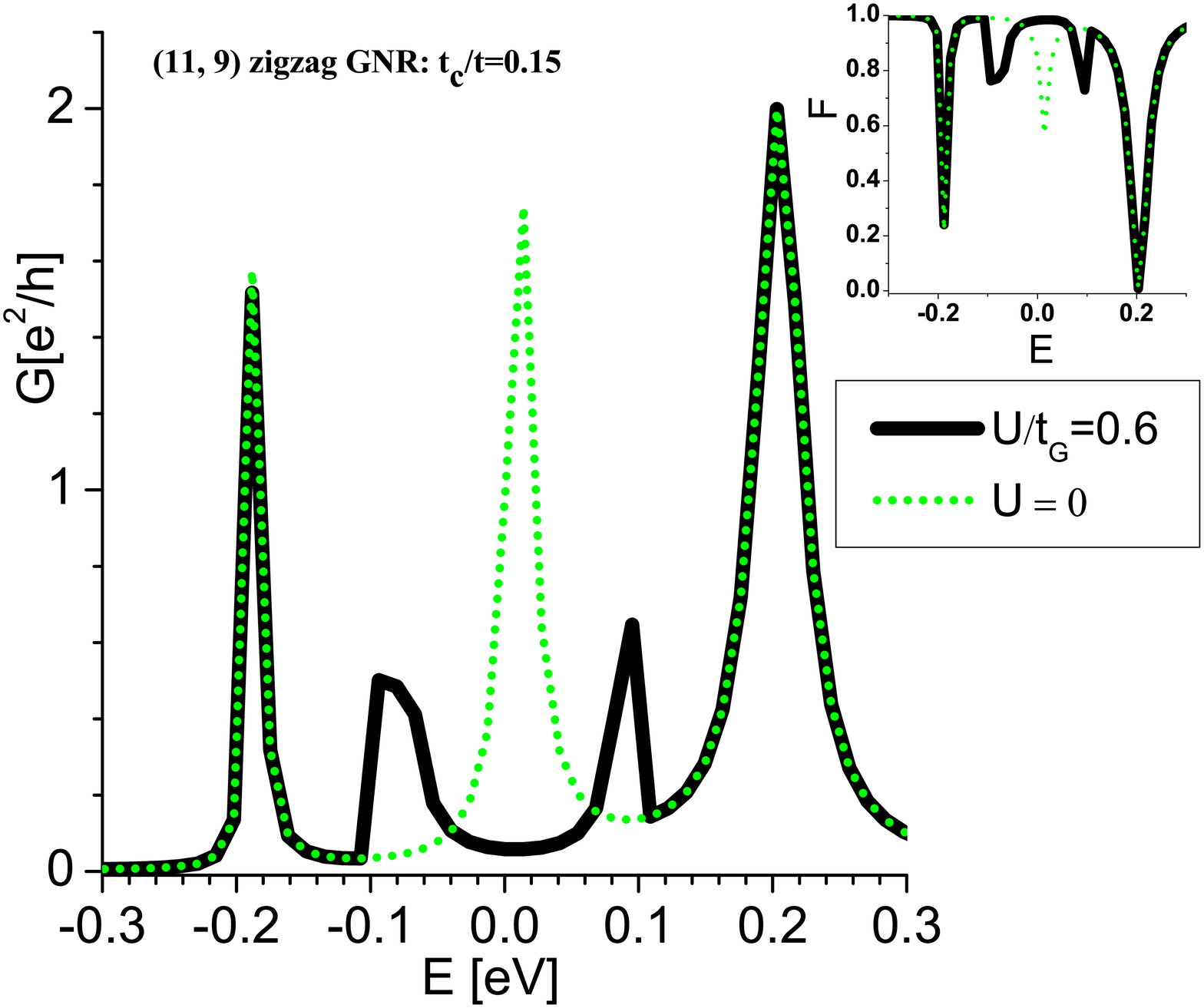}
 \caption{\small (Color online) Conductance and the shot noise
 Fano factor for the (11, 9) GNR. Close to E=0 the plots for
 finite U and U=0 differ greatly, otherwise they coincide with each other.} \label{GiF_tc.15}
   \end{figure}
   \end{center}

\begin{center}
  \begin{figure}[h]
  \includegraphics[width=10cm,height=7cm,angle=0]{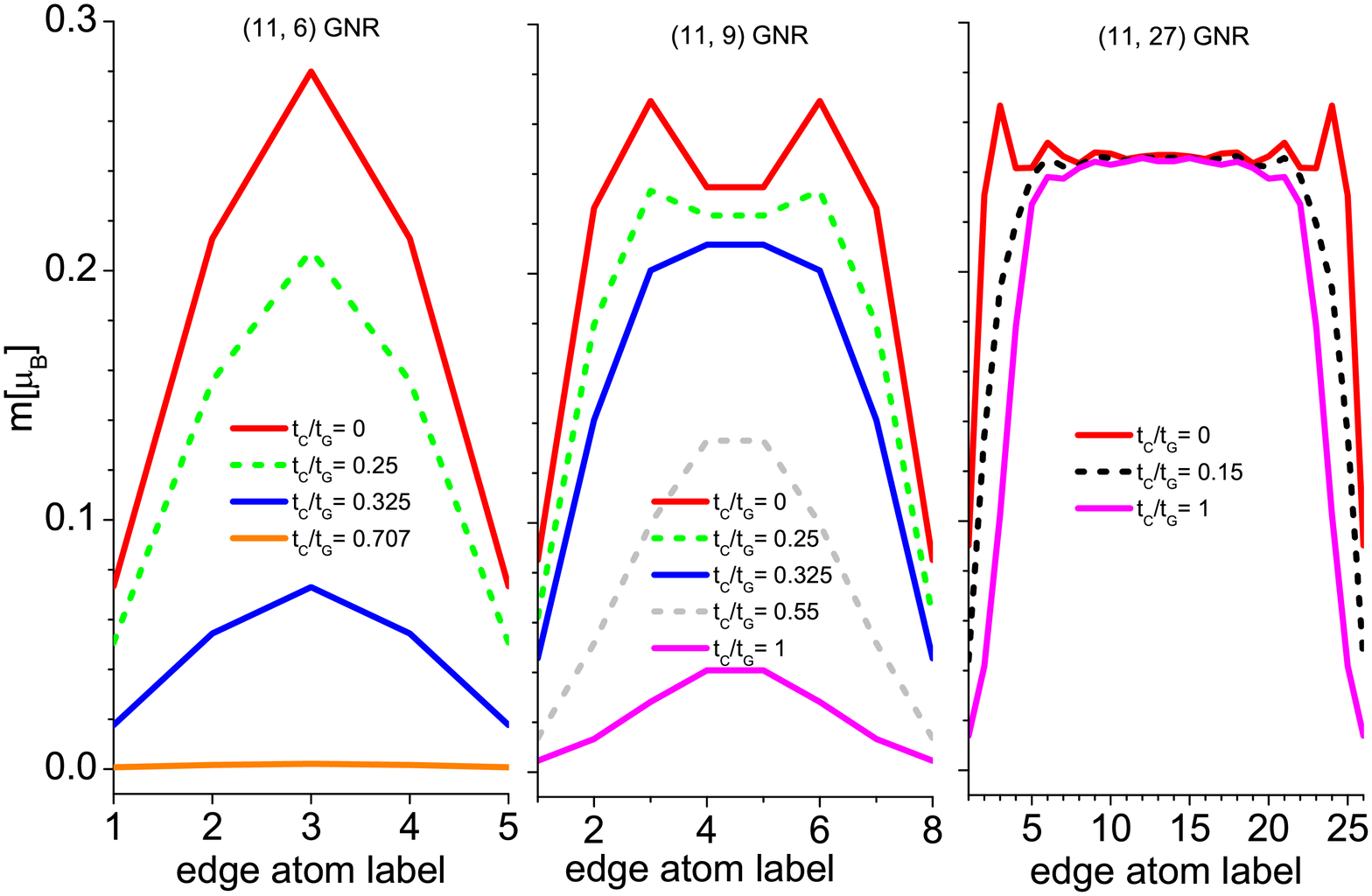}
   \caption{\small (Color online) Edge magnetization profiles for
  3 different aspect ratio GNRs, indicated $t_c$ parameters, and $E_F=0$.
  Magnetic moments at the other zigzag edge are oppositely oriented. } \label{E03M}
   \end{figure}
   \end{center}

The presence of the contacts results in a slight shift of the
Fermi energy with respect to the unperturbed case of $E_F=0$. In
order to gain a deeper insight into the doping problem, the
magnetic profiles along the zigzag edges  have been first determined for
the Fermi energy $E_F=0$ (as for the free standing GNR with $t_c=0$), Fig.~\ref{E03M},
and next for $E_F$ corresponding to the charge neutrality point
(CNP).

\begin{center}
  \begin{figure}[b]

\includegraphics[width=10cm,height=7cm,angle=0]{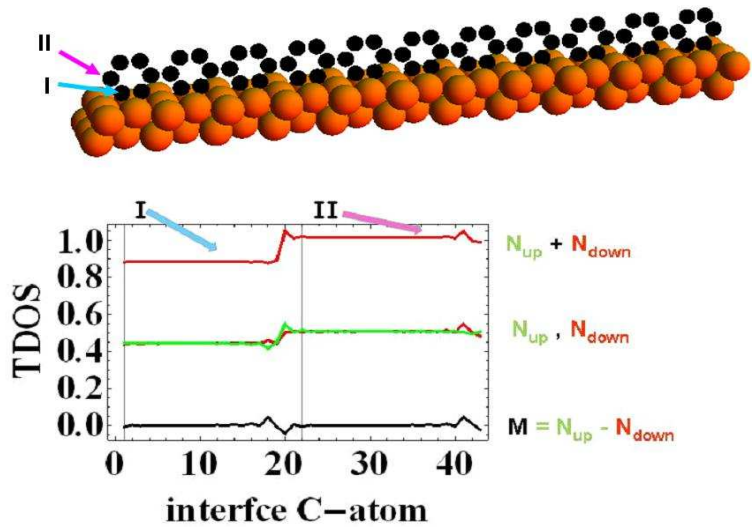}
 \caption{\small (Color online) The profiles of the number of electrons in the (11, 27) GNR with $t_c/t_G=0.15$. Region I corresponds to the interface carbon atom line (nearest to the contact), whereas the region II is the next nearest line of atoms. $N_{up}$ ($N_{down}$) stands for the number of spin up  (down) electrons. Charge depletion in the
 region I is readily seen.} \label{interface}
   \end{figure}
   \end{center}

Although, in general, unintentional doping coming from electrodes
depends on the difference in respective work functions (WFs) of
the electrodes and the GNR, it turns out that the p-type doping is
expected in the case of high-WF metals that couple weakly to GNR
(e.g. Au \cite{Giovannetti_PRL2008, Li_PRL2013}). As shown in
Fig.~ \ref{interface} the present model fits qualitatively to this
scenario.

\vspace{0.5cm}    
\begin{center}
  \begin{figure}[t]
\includegraphics[width=10cm,height=7cm,angle=0]{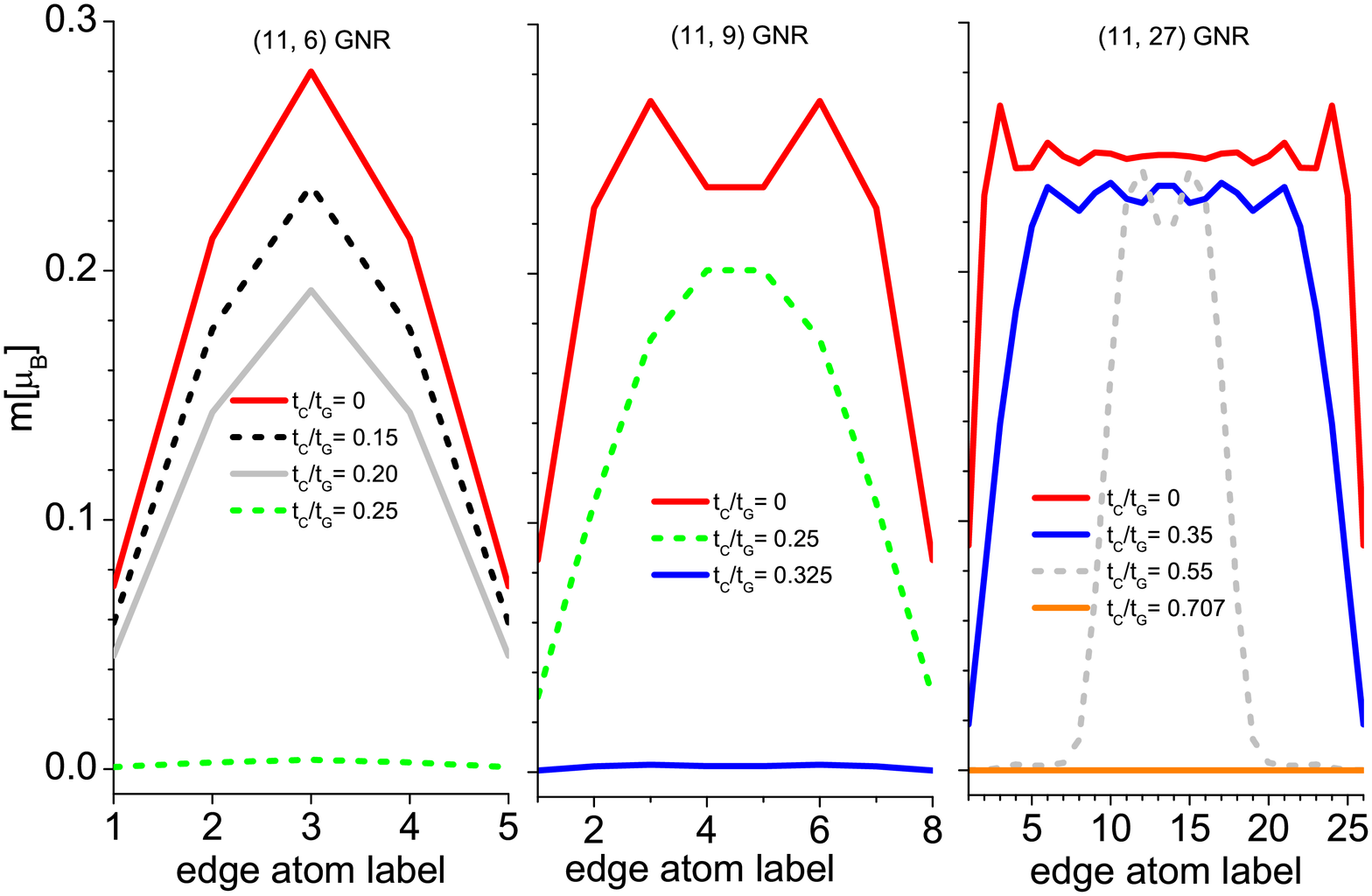}
 \caption{\small (Color online) As Fig.~\ref{E03M}
 but for Fermi energies equal to the charge neutrality points.} \label{CNPM3}
   \end{figure}
   \end{center}

In Fig.~\ref{CNPM3}, in turn, the $E_F$ correspond to the actual
CNPs. It is readily seen that both in Fig.~\ref{E03M} and
Fig.~\ref{CNPM3} the increasing strength of the $t_c$ coupling
suppresses edge magnetic moments. The effect is the most
pronounced when the setups are wide and short (big A). However, in
the case of $E_F=CNP$ it is possible to quench completely all the
magnetic moments even for the (11, 27) GNR (with the aspect ratio
clearly less than one), provided that $t_c \sim 0.7 t_G$.

The present findings agree qualitatively with Ref.
\cite{Li_PRL2013}, where edge states and edge magnetism in
graphene nanoribbons were studied by means of
  an {\it ab initio} method. Although those studies concerned  the substrate effects instead of the contact ones, they led to conclusions which bear some
resemblance to the present ones. First of all, the edge magnetism
was shown there  to disappear in the case of closed-packed
surfaces of the substrates with a strong graphene/substrate
coupling (Cu and Ag), in contrast to the Au substrate of weaker
coupling. These metallic substrate effects may be related to the
present results in terms of our $t_c$ parameter, which for the Cu
and Ag substrates is clearly higher than for the Au one, due to
differences in the respective carbon/substrate-atom distances.
Remarkably, typical edge magnetic moments (of roughly 0.2 - 0.3
$\mu_B$  per C atom) found here do also agree with those reported
in Ref. \cite{Li_PRL2013}. So, despite the fact that the present
approach is qualitative rather than quantitative, it provides some
indirect insight into the chemical nature of the contacts. Indeed
the phenomenological parameters $t_c$, which in fact depend on
orbital hybridization and a bond length at the interface, can be
related to material specific \textit{ab initio} results.

\section{Conclusions}

Summarizing, in accordance with common knowledge it has been
confirmed here that non-contacted (free standing) GNRs reveal edge
magnetism in the zigzag outermost carbon atoms. Remarkably, the
magnitude of the edge magnetic moments is sensitive to the
GNR/contact coupling strength and the separation of the given edge
atom from the interface. In the case of high aspect ratio GNRs the
zigzag edge magnetic moments may be severely reduced or even
completely quenched. The present findings concerning the
conductance and the Fano factor indicate that magnetic moments can
also be suppressed by applying a gate voltage that brings the
system to the state far enough away from the charge neutrality
point.

\section{Acknowledgments}
This project was supported by the Polish National Science Centre
from funds awarded through the decision No.
DEC-2013/10/M/ST3/00488.

\end{document}